\documentclass{aa}
\usepackage{graphics}
\usepackage{psfig}
\begin{document}
\def \arcm{\hbox{$^\prime$}}
\def \arcs{\hbox{\arcm\hskip -0.1em\arcm$\;$}}
%\newenvironment{decl}
%  {\par\small\addvspace{1.9ex plus 1ex}
%    \vskip -\parskip
%    \noindent\hspace{-\leftmargini}
%    \begin{tabular}{|l|}\hline\ignorespaces}
%  {\\\hline\end{tabular}\nobreak
%    \vspace{2.2ex}\vskip -\parskip}

\title{Deep optical observations of the fields of
two nearby millisecond pulsars
%PSR J1024-0719 and
%PSR J1744-1134
with the VLT
\thanks{Based on
observations with the ESO 8.2m VLT-Antu (UT1) Telescope under
Program 66.D-0448 and 67.D-0538}}

\author{ F. K. Sutaria \inst{1}
     \and A. Ray \inst{2}
     \and A. Reisenegger \inst{3}
     \and G. Hertling \inst{3}
     \and H. Quintana \inst{3}
     \and D. Minniti \inst{3}
    }

%\offprints{A.Reisenegger,areisene@astro.puc.cl}
%\email{F.K.Sutaria@open.ac.uk}}
%\begin{decl}
%\verb|\offprints{A. Reisenegger}|\\
%\verb|\email{areisene@astro.puc.cl}|
%\end{decl}
\institute{Dept. of Physics and Astronomy, The Open University,
Milton Keynes, U.K.
%                                            \mail{F.K.Sutaria@open.ac.uk}
           \and Tata Institute of Fundamental Research, Mumbai, India.
           \and Dept. of Astronomy and Astrophysics, Pontificia Universidad Cat\'olica de Chile, 
           Santiago,
           Chile. \mail{areisene@astro.puc.cl}}

\authorrunning{F. K. Sutaria et al.}
\titlerunning{ Optical observations of two nearby millisecond pulsars with the VLT}
\date{Received ; Accepted }
\maketitle
\abstract{We report on deep VLT observations of the fields of
two nearby, isolated millisecond pulsars PSR J1744-1134 and PSR J1024-0719.
Both objects are old neutron
stars with characteristic age $\tau \ge 10^9$ yr and have relatively high
spin-down flux. They have been detected earlier as
X-ray sources by ROSAT HRI observations and were considered good
candidates for non-thermal emission in the optical bands. Our
observations set an upper limit of $B=26.9$, $V=26.3$ and $R=26.0$
for PSR J1744-1134. In the case of PSR J1024-0719, we find two
faint objects near the radio position of the pulsar.
%with magnitudes $U=22.56$, $V=23.42$, $R=22.83$ and $I=22.7$.
Using multi-band photometry from the VLT and spectroscopy carried
out with the Magellan I telescope, we discuss the nature of the
brighter object and the possibility of the fainter one being the
optical counterpart of PSR J1024-0719. We consider the
implications of our findings for both pulsars in the context of
theoretical models of high-energy emission from old pulsars.

\keywords{Stars: neutron; Pulsars: general; Pulsars: individual:
PSR J1024-0719, PSR J1744-1134; Radiation mechanisms: non-thermal;
Radiation mechanisms: thermal}}

\section{Introduction}
\label{sec: Intro}

Optical pulses from the Crab pulsar were detected by Cocke, Disney
and Taylor (1969) within a year of the discovery of the pulsar in
the radio band (\cite{Staelin68}) and the inauguration of the
field of pulsars by S. Jocelyn Bell and collaborators
(\cite{Hew68}). Since then, the number of pulsars discovered in
the radio band has gone up to nearly 1400 (ATNF pulsar catalogue),
but the number of pulsed counterparts detected in the high-energy bands
remains small. To date, in the optical bands, these remain 5, 
(\cite{Shearer02}), and in the X-ray band 22, including  
6 millisecond pulsars (MSP) (\cite{Beck02}). Multi-wavelength detection
of rotation-powered neutron stars provides observational input to
the nature of the underlying mechanism of pulsar radiation,
whether it is of thermal origin or due to radiation from
accelerated particles in its magnetosphere (see e.g. Pavlov,
Stringfellow, \& Cordova 1996).

Millisecond pulsars (MSP) are a special class, not only because of
their short spin periods, but also because of their smaller
surface magnetic field, old age, and evolutionary history of
spin-up due to accretion of mass and angular momentum from a
binary companion (\cite{Alp82}). In the process, the star may have
been resurrected from a state of hibernation into one of
detectable pulsed radio emission. They represent a considerable
extension of the parameter space of classical pulsars, so
different radiation mechanisms may well be active in their
magnetospheres. Their old age means that they have almost
certainly radiated away any ``fossil'' heat from their original
collapse or from their accretion epoch. Thus, any thermal emission
must be due to external reheating from the magnetosphere or to
internal dissipation mechanisms, most likely related to the star's
slowing rotation rate (see, e.g., \cite{Che92}, \cite{Rei95},
\cite{Lar99}, and references therein).

While MSP with binary companions are interesting
in their own right from the evolutionary angle, radiation from the
companions, even if they are low-mass old white dwarfs,
``contaminates'' the faint high-energy radiation intrinsic to the
pulsar itself. This is especially true of the optical and
ultraviolet regions. It is therefore of interest to concentrate on
the isolated pulsars to study their radiative properties in these
bands. Isolated MSPs are relatively rare objects --- till mid-2002,
only 10 have been detected in the Galactic place. By contrast, the
number of galactic MSPs (outside globular clusters) totals $\sim 57$. 
None of the optically detected pulsars is an isolated MSP, which
might at least partly be due to a lack of sufficiently deep exposures.
The advent of the VLT class of telescopes is therefore a crucial
technological development for these faint objects.
%In Table (\ref{tab1: optical_status}) we give a list of {\it isolated, X-ray
%detected, millisecond} pulsars and their optical {\it observation status}.

The pulsed high-energy radiation from a rotation powered
(isolated) pulsar is a combination of differing amounts of three
spectral components: 1) power law emission resulting from
particles accelerated in the magnetosphere, 2) a soft black-body
component from the surface of a cooling neutron star, and 3) a
hard thermal component associated with heated polar caps bombarded
by energetic particles.  Measurements of the radiative fluxes
and spectra in different bands can constrain the relative weights of
these components.

In this paper, we present the results of deep VLT observations of
two southern, isolated millisecond pulsars, PSR J1024-0719 and PSR
J1744-1134, which are among the closest MSPs
discovered and which have also been detected in the X-ray band by
ROSAT. Identification of candidate objects in the optical band
forms the first step towards searches for optical pulsations from
a pulsar. These pulsars have some of the highest spin-down flux
($\dot E_{rot} / 4 \pi d^2 = I \Omega \dot\Omega / 4 \pi d^2$)
among millisecond pulsars, an indicator of their propensity to
produce high-energy radiation. Because of their proximity as well
as low extinction in their direction,
they are among the best candidates for optical detection. In
Sect.  \ref{sec: prevobs}, we discuss
what is known about these two pulsars in the X-ray and radio
bands. We next present, in Sect. \ref{sec: obs}, the VLT service mode
observations and their data analysis. In Sect. \ref{sec: results},
we discuss the results obtained from the VLT data. We supplement
our discussion of a star near PSR J1024-0719 with a spectroscopic
observation with the Magellan I Baade 6.5 m telescope at Las
Campanas Observatory. In Sect. \ref{sec: disc}, we discuss the
multi-band optical results in the context of high-energy and radio
emission from the pulsars and compare these with those from slower
pulsars, and theoretical models of pulsar radiation. We give our
conclusions in Sect. \ref{sec: conclusn}.

\section{Radio and X-ray properties of the two pulsars}
\label{sec: prevobs}

Both PSR J1744-1134 and PSR J1024-0719 were first detected in the
radio bands as isolated MSP with periods
$P=4.07$ and 5.16 ms, respectively (\cite{Bailes97}).  In Tab.
\ref{tab: PSR} we quote some characteristics of these objects
relevant to our observations and their interpretation.  The
astrometric data and the proper motions were determined by radio
timing observations (\cite{Tos99}).

The radio pulse profile of PSR J1744-1134 is narrow and sharply
peaked, with a duty cycle of $\sim 10 \%$, while PSR J1024-0719
exhibits a broad, multiply peaked profile with a duty cycle of
$\sim 50 \%$.
The radio and X-ray emission are most likely based on different
physical mechanisms: curvature radiation from $e^{\pm}$ pairs near
the NS surface (\cite{Ruderman75}), and synchrotron radiation from
energetic particles in the outer gap (\cite{Cheng86}),
respectively. However, at least in the case of PSR B1821-24
(another MSP), RXTE observations (\cite{Rots97}) show that the
leading edge of the most energetic pulse coincides with the radio
pulse profile at 800 MHz, possibly implying a common site for the
origin of the two pulses.

Nevertheless, the two pulsars are similar in that the spin-down
ages for both objects are $\tau \equiv P/(2\dot P) > 10^{9}$ yr
(Tab.  \ref{tab: PSR}).
Both pulsars are in the age bracket where
non-thermal magnetospheric emission can be expected
to dominate over any thermal effects,
unless strong re-heating occurs.
Some uncertainty exists over the distance estimate to the two
pulsars. An upper limit to the distance of PSR J1744-1134, based
on the assumption that the period derivative is entirely due to
the transverse motion of the pulsar (Shklovskii effect), is 1910
pc (\cite{Tos99}). Based on parallax measurements of PSR
J1744-1134, \cite{Tos99_2} estimate the distance $d$ as
$357^{+43}_{-35}$ pc.  This number decreases to 166 pc
(\cite{Bailes97}) if $d$ is estimated using the measured value of
the dispersion measure (DM) and values of electron densities from
the \cite{Tay93} model. However, it is to be noted that, while
this model adequately describes the properties of interstellar
medium (ISM) at $d>1$ kpc, it does not account properly for
fluctuations in the local ISM. For our purposes, we use a distance
estimate of 0.357 kpc for PSR J1744-1134. In the case of PSR
J1024-0719, an upper limit of $d = 0.226$ kpc is obtained from the
Shklovskii effect, while a $d=0.35$ kpc is obtained by using DM
and the Taylor and Cordes model (\cite{Tos99}).
We shall adopt a value $d=0.2$ kpc for PSR J1024-0719
in this paper, unless specified otherwise.

 The corresponding X-ray counterparts for both pulsars were discovered
by the ROSAT HRI (\cite{Beck99}; hereafter BT99). Because both
pulsars are quite faint in this band, no timing observations could
be done, and the counterparts were identified by their proximity
to the radio positions. Assuming
%a column density $N_H = 2 \times 10^{20}$ cm$^{-2}$, and
a photon spectral index $\alpha=2$, these authors find that the
unabsorbed X-ray luminosity of PSR J1744-1134 (scaled to our
adopted distance) is $L_X = 4 \times 10^{29} (d/0.36 {\rm
kpc})^2$, and that of PSR J1024-0719 is $ L_X = 1 \times 10^{29}
(d/0.2 {\rm kpc})^2$.
The X-ray to spin-down luminosity ratios $L_X/\dot{E}_{rot}$ are
then $2\times 10^{-4}$ and $5\times 10^{-5}$, respectively,
somewhat below the general relation that \cite{Beck97} found
statistically among high-energy pulsars, $L_X/\dot{E}_{rot}\sim
10^{-3}$.

\begin{table}[h]
\caption{ Characteristics of PSRs J1024-0719 and J1744-1134 (BT99,
\cite{Tos99}, \cite{Tos99_2})} \label{tab: PSR} \vskip 0.5 true
cm\begin{tabular}{ccc} \hline \hline
 Property  &  PSR J1024-0719 &  PSR J1744-1134 \\
\hline Period\,($P$)\,[ms] & 5.16 & 4.07 \\
 $\dot P$ $[10^{-20}{\rm s\;s}^{-1}]$ & 1.873(5) &  0.89405(9) \\
 $\dot E_{rot}$ $[{\rm erg\; s}^{-1}]$ & $5.25\times 10^{33}$   & $1.90\times 10^{33}$ \\
%$ $ &     &   \\
 $d$ [kpc] &  0.200 &  0.357 \\ %(0.26 from Becker and Pavlov Tab. 2)
 $\tau$ [$ 10^9$ yr] & 27 & 9.1 \\
 $B$ [$ 10^8$ G] &   1.3  &  1.7  \\
 R.A. (J2000)  & 10$^h$24$^m$38\fs7040(1) & 17$^h$44$^m$29\fs390963(5) \\
 Dec. (J2000)  & -07$^{\circ}$19\arcm18\farcs849(3) & -11$^{\circ}$34\arcm54\farcs5746(5) \\
 P. M. (R.A.) & -41(2) &  18.72(6) \\
 $[{\rm mas\; yr}^{-1}]$   &   &  \\
 P. M. (Dec.)  &  -70(3) &  -9.5(4) \\
 $[{\rm mas\; yr}^{-1}]$  &   &  \\
 Epoch (MJD) & 50456.0  &  50434.0 \\
% X-ray counterpart & -- &  un-pulsed  &  un-pulsed \\
 $L_X$(0.2-2.4\,keV)  & $1\times10^{29}(d/200{\rm pc})^2$ &
 $4\times10^{29}(d/360{\rm pc})^2$ \\
 $[{\rm erg\; s}^{-1}]$ &  & \\
\hline
\end{tabular}
\end{table}

\section{The VLT observations}
\label{sec: obs}

PSR J1744-1134 was observed by the ESO 8.2m Very Large Telescope
Antu (VLT-UT1) with the FORS1 CCD in the narrow field imaging mode
in the Bessel B, V and R bands on four nights during April 2001
(MJD 52017-52027). PSR J1024-0719 was observed by the same
instruments/mode in the U, V, R and I bands within less than one
hour in March 2001 (MJD 51996). Both sets of observations were
carried out in the service mode. A brief summary of the
observations is provided in Tab. \ref{tab: obs}.

\subsection{The observation}

The photometric analysis for both objects was done using the MIDAS
pipeline processed data. The pipeline processing ensures that each
frame is checked for over-exposure and suitably bias-subtracted
and flat-fielded, using the master bias and flats generated from
calibrations observations made on each night. Finally, the prescan
and over-scan regions are removed from the science images used in
this analysis.

Our initial examination of the DSS plates in the vicinity of PSR
J1744-1134 showed no bright objects. However, the VLT images show
a bright star with $V=21.3$ about $2\farcs0$ from the position of
the pulsar. Thus, in order to strike a balance between deep
exposure and good seeing, of the twenty-three ten-minute exposures
with Bessel B, we rejected the three exposures with the worst
seeing taken on 2001 April 21, thus reducing the available
photometry time to 3.33 hr. The available exposure time in each of
the V and R filters was 48 min. The median seeing in the B, V and
R filters was $0\farcs725$, $1\farcs095$ and $0\farcs795$
respectively.

For PSR J1024-0719, the available exposure time was 13.33 min in
U, 6 min in both V and I, and 9 min in R (see table \ref{tab:
obs}). The median seeing in the U, V, R and I frames was
$0\farcs63$, $0\farcs97$, $0\farcs74$ and $0\farcs71$
respectively.

\begin{table}[h]
\caption{ Observation summary for VLT FORS1 CCD photometry}
\label{tab: obs}
\begin{tabular}{ccccc}
\hline
\hline
 PSR &  Filter & Exposure &  Mean &  Mean  \\
     &         & time &  air  & Seeing \\
        & (Bessel) & [s] & mass  &  [$\arcs$]\\
\hline
 J1744-1134    & B & $ 20 \times 600  $ &  1.04  &  0.725 \\
               & V & $  2 \times 540  $ &  1.06  &  1.095 \\
               &   & $ +3 \times 600  $ &        &        \\
               & R & $  2 \times 540  $ &  1.08  &  0.795 \\
               &   & $ +3 \times 600  $ &        &        \\
\hline
 J1024-0719    &  U & $ 1 \times 900  $ &  1.2   &  0.63 \\
               &  V & $ 3 \times 120  $ &  1.3   &  0.97 \\
               &  R & $ 3 \times 180  $ &  1.2   &  0.74 \\
               &  I & $ 3 \times 120  $ &  1.2   &  0.71 \\
\hline
\end{tabular}
\end{table}

\subsection{Data reduction}
\label{sec: datared} The image processing was done using the IRAF
software. With the exception of the single image in the U-band,
the reduced images in each filter were aligned and median-combined
using a cosmic-ray rejection algorithm. Because in the case of
both PSR J1024-0719  and PSR J1744-1134, there is a $V \sim 20-21$
star present within $\sim 1\farcs5 - 2\arcs$ of the pulsar position,
photometric analysis of the combined images was carried out using
PSF-subtraction photometry (\cite{Massey}). In order to ensure
that wings of the PSF of nearby stars are subtracted out cleanly,
we constructed a PSF for each frame, using only the brightest
unsaturated stars, which were of magnitude a little brighter or
close to that of the ``contaminating star''. In general, we find
that the the PSF of the FORS1-CCD is position-dependent and is
best fit by the IRAF moffat25 or penny1 functions with 2nd order
variability in X and Y. Aperture correction was done on the CCD
magnitudes, and the final magnitudes of all objects observed in
the vicinity of the pulsar's radio position were determined by
comparison with standard stars. The standard stars used were from
\cite{Land92} field SA109 for PSR J1024-0719, and fields SA109,
SA110, and MARKA for PSR J1744-1134, observed on the same nights.

\subsection{Astrometry}
\label{sec: astrometry}

Correcting for the proper motion, the position of PSR J1744-1134
at the epoch of the the VLT observations was RA $= 17^h 44^m
29\fs39638(3)$, Dec $= -11^{\circ} 34\arcm 54\farcs616(2)$, and
that of PSR J1024-0719 was RA $= 10^h 24^m 38\fs6925(6)$, Dec $=
-07^{\circ} 19\arcm 19\farcs14(1)$.

Astrometric corrections to the observed CCD positions were carried
out by comparison with both USNO-A2 catalog of astrometric
standards (\cite{Mon98}) and the HST Guide Star II (GSCII)
catalog. We report the results based on the more recent GSCII
catalog, where the epoch of observation was from 1983-1986. The
presence of nearby bright objects made it imperative to observe in
the high-resolution mode. However, this reduced the number of
GSCII astrometric standards in the field of view (FOV) of PSR
J1024-0719 to only 6. The crowded field PSR J1744-1134 provided 14
astrometric standards within $1\farcm5$ of the pulsar position.
The pixel coordinates of the reference stars were obtained from
our photometric analysis, and they were converted to RA and Dec
using the ASTROM package supplied with the STARLINK software.

For the field of PSR J1024-0719, the rms error in the fitting and
transformations was $0\farcs12$ in RA and $0\farcs16$ in Dec.  The
average error in the positions of the relevant stars in the GSCII
catalog is $0\farcs39$ in both directions. Thus, incorporating the
error in the measurement of the proper motion, we estimate the
total astrometric error in the observation of this field at
$0\farcs42$ in RA and $0\farcs43$ in Dec. For PSR J1744-1134, the
error in the astrometry of the GSCII stars is $0\farcs28$ in RA
and $0\farcs30$ in Dec, and the error in the fitting and
transformations is $0\farcs16$ and $0\farcs14$, respectively. In
this case, the total astrometric error in the position is
$0\farcs34$ and $0\farcs33$ in RA and Dec, respectively. We note
that astrometry relative to the USNO-A2 catalog provides similar
results, though the errors are slightly larger.

Since we use the radio pulsar positions to search for their
optical counterparts, we need to consider the accuracy of the {\it
radio timing} position with respect to the position in the optical
frame. The GSCII catalog was calibrated using the Hipparcos and
Tycho frames of reference, which are tied to the International
Celestial Reference Frame (ICRF, tied to the radio Very Long
Baseline Interferometry, \cite{Ma98}).

An estimate of the absolute accuracy of pulsar radio timing
positions was made by \cite{Fom92} (and \cite{Fom97}) through
comparing them with positions determined by interferometry. In
general, they found the RA and Dec position offsets derived from
timing and interferometric positions to be a function of the
position of the target in the sky. Fomalont et al. (1992) state
that, while there are no significant offsets between the
coordinate frames in a global comparison of VLA positions and
timing positions, there is a mean scatter of about $0\farcs8$ in
each coordinate (RA and Dec).

If we add this to the above astrometric errors, then for both
pulsars the total astrometric error in both RA and Dec would be
$\approx 0\farcs9$, giving an error circle of radius $\approx
1\farcs3$. Therefore, any optical star-like emission within about
$1\farcs5$ of the radio timing position should be taken with
interest for further scrutiny, especially for time resolved
photometry.

\section{Results}
\label{sec: results}

\begin{figure}[h]
\caption{VLT FORS1 V-band field of view $ 40\arcs \times 40\arcs $
centred about the radio timing position of PSR J1744-1139 (marked
by the cross). North is to the top, and East to the left. 
\label{fig: fov1744-11}} 
\vskip 0.8cm
%\resizebox*{\hsize}{!}{\includegraphics{1744_finding_chart.ps}}
\resizebox*{\hsize}{!}{\includegraphics{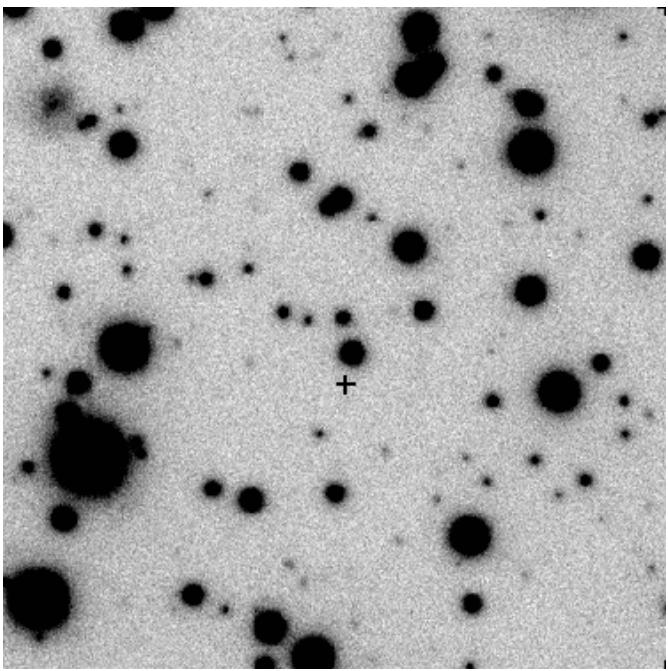}}
\end{figure}

The broadband magnitudes in multiple optical bands and astrometric
positions of objects within $2\farcs0$ of the radio timing
positions of our two target millisecond pulsars are reported in
Tab. \ref{tab: mag}. We do not detect any object near PSR
J1744-1134. An image of the field of view of this pulsar appears
in Fig. \ref{fig: fov1744-11}.

\begin{figure}
\caption{Images (a) and (b) are the Field of View (FOV) of PSR
J1744-1134 in V, before and after PSF-subtraction, respectively.
The ``error circle'' of $1 \arcs$ is centered at the radio
position of the PSR. Images (c) and (d) show the FOV of PSR
J1024-0719 in U, before and after PSF-subtraction respectively.
The position of the pulsar within the $1 \arcs$ ``error circle''
is marked by the small white arrow. The blue object 1024-Fnt is
encircled within the smaller region, just north of 1024-Br.
\label{fig: Psf_Image}} 
\vskip 0.3cm
%\resizebox*{\hsize}{!}{\includegraphics{Psf-Image.ps}}
\resizebox*{\hsize}{!}{\includegraphics{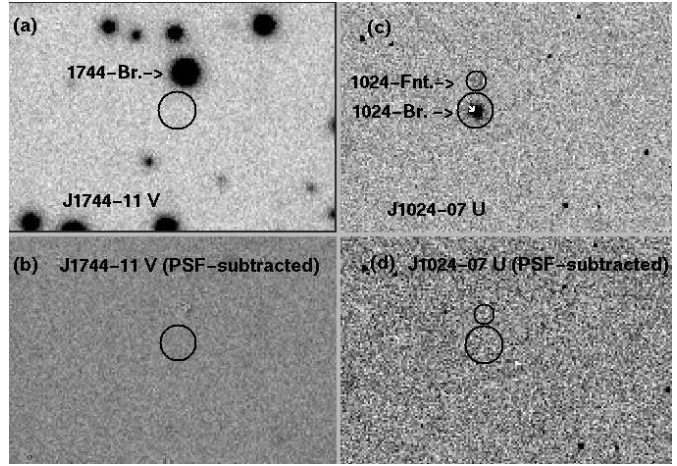}}
\end{figure}

Because of the presence of a $V=21.34$ star close to the position
of PSR J1744-1134, the limiting magnitude for the pulsar
counterpart was estimated by introducing artificial stars at the
pulsar position in each filter, by using the IRAF task
{\sf addstar}, and repeating PSF-subtraction photometry (see Fig.
\ref{fig: Psf_Image}). Thus, the limiting magnitude for
3.5-$\sigma$ detection by the automatic star-finding algorithm
{\sf daophot} are $B=26.9$, $V=26.3$, and $R=26.0$. We note that,
from exposure-time considerations, the faintest object that could
be detected in these frames would have $B=28.0$, $V=27.25$,
$R=26.8$ for a 4-$\sigma$ detection over the entire PSF.

For PSR J1024-0719,
%the corresponding limits at the nominal radio
%timing position centroid are  given as ``upper lim.'' in Tab.
%\ref{tab: mag}. We do however
we find two objects within $1\farcs5$ of the radio position of the
pulsar (see Fig. \ref{fig: Psf_Image}). The closer one, 1024-Br,
only $\sim 0\farcs2$ away,
%(see Fig. \ref{fig: UBESS}).
has magnitudes $U= 22.11 \pm 0.03$, $V=19.82 \pm 0.01 $, $R= 18.89
\pm 0.01 $, and $I= 18.17 \pm 0.01$.
%PSF-subtraction removes this
%object cleanly in the U frame, implying an exposure-time based
%upper limit of $U=24.5$ (3$\sigma$ detection). The magnitudes in
%the V, I and R frames are given in the same Table.
%We mention here that both USNO and 2MASS catalogue report
%an object is present near  1024-Br and give additional
%data: $R =19.1$ and $J = 17.17$ and $H = 16.12$ respectively.
The other, substantially fainter object, 1024-Fnt, $\sim
1\farcs48$ north of the radio pulsar, has magnitudes $U=23.8 \pm
0.1$, $ V= 24.9 \pm 0.1 $, $R=24.4 \pm 0.1 $ and $I=24.2 \pm 0.2
$. The reported magnitude errors for both objects include the
intrinsic error in the Landolt catalogue and the error in
photometry. The angular separation between the bright and faint
objects is $1\farcs67$.

An object is reported close to the position of 1024-Br in various
optical and infrared catalogs. The USNO-A2 catalog (Monet et al.
1998) gives magnitudes $B=19.7$ and $R=19.1$, DENIS (DENIS
Consortium, 1998) reports $I=18.0$, and 2MASS (IPAC/UMass, 2000)
gives $J=17.2$, $H=16.1$, and $K=16.5$). Applying astrometric
corrections based on coincident stars within our field, we find
the DENIS and 2MASS positions to agree with our determination for
1024-Br to within $0\farcs2$, whereas the USNO-A2 position is off
by $+1\farcs2$ in RA and $+2\farcs0$ in Dec. The similarity in
positions and magnitudes makes it almost unavoidable that the
source is the same in all cases. The discrepancy in the USNO-A2
position (taken in 1953, whereas all other measurements are from
1995-2001) may be either an error or an indication of a proper
motion of $-25\rm \;mas\; yr^{-1}$ in RA and $-40\; \rm mas \;
yr^{-1}$ in Dec, almost identical in direction, but about $40\%$
lower in magnitude, compared to the radio-timing-determined proper
motion of PSR J1024-0719.

\begin{table}[h]
\caption{ Positions
and magnitudes of optical point sources closest to the radio positions of
two MSPs.
The limiting magnitudes are also provided for $3.5 \sigma$
detection at the central pixel in the exposure times used here for PSR J1744-1134}
\label{tab: mag}
\begin{tabular}{cccc}
\hline
\hline
 Object &  Magnitude & \multicolumn{2}{c}{ PSR pos. - Obj. pos.}\\
       &           &  $ \delta$ RA &  $ \delta$ Dec \\
%       &           &   (\arcs)     &   (\arcs) \\
\hline
\hline
 J1744-1134 &  $B\geq$26.9, $V\geq$26.3, &  0\farcs0  &  0\farcs0 \\
 (upper lim.) &  $R\geq$26.0        &       &      \\
1744-Br &  $B=18.86$, $V=21.34$, & $+0\farcs4$ & $-2\farcs0$ \\
% 1744-Br &  $B=18.86$, $V=21.34$, & $+0\farcs426$ & $-1\farcs997$ \\
          &  $R=19.91$           &       &       \\
\hline
% J1024-0719 &  $U\geq$24.5, $V\geq$26.05,   & 0.0 & 0.0 \\
% (upper lim.) & $R\geq$25.9, $I\geq$24.8    &     &     \\
 1024-Br &  $U=22.11$, $V=19.82$,  & $+0\farcs1$   & $-0\farcs2$  \\
% 1024-Br &  $U=22.11$, $V=19.82$,  & $+0\farcs110$   & $-0\farcs183$  \\
          &  $R=18.89$, $I=18.17$   &      &      \\
1024-Fnt &  $U=23.8$,  $V=24.9$,   & $ -0\farcs2$  & $+1\farcs5$ \\
%1024-Fnt &  $U=23.8$,  $V=24.9$,   & $ -0\farcs177$  & $+1\farcs466$ \\
          &  $R=24.4$,  $I=24.2$    &      &     \\
\hline
\end{tabular}
\end{table}

\subsection{Extinction-corrected magnitudes and multi-waveband fluxes}
\label{sec: multiband}

Based on the dispersion measures of the two pulsars, BT99 %\cite{Beck99}
obtained hydrogen column densities in the directions of PSR
J1024-0719 and PSR J1744-1134 as $N_H = 2 \times 10^{20}\; \rm
cm^{-2}$ and $N_H = 1 \times 10^{20}\; \rm cm^{-2}$ respectively.
In the ROSAT $0.1-2.4$ keV band, their X-ray fluxes were both $f_X
\sim 2 \times 10^{-14} \rm erg\; s^{-1} cm^{-2}$. Using the
scaling of \cite{dB87} between visual extinction and hydrogen
column density, $A_V = N_H /(1.79\times 10^{21})$, we obtain $A_V
= 0.11$ for PSR J1024-0719. The wavelength-dependent continuum
absorption in the main photometric bands (\cite{All00}) then
gives, for the same pulsar, $A_U = 1.58 A_V = 0.17$, $A_R = 0.749
A_V = 0.08$, and $A_I = 0.479 A_V = 0.05$.

\begin{figure}
\caption{Optical photometry of pulsars. The top set ($\triangle$) is
for the Crab pulsar with flux densities scaled down by a factor of 100
to fit in the graph. It also includes a measurement in the UV band at
$\lambda\lambda 2400 A$. Next is the LMC pulsar PSR 0540-69
($\diamond$), followed by the Vela pulsar PSR 0833-45 (squares).
Finally, at the bottom are extinction-corrected flux densities
of the object 1024-Fnt ($\times$). \label{fig: multibandpsr}}
\vskip 0.7cm
%\resizebox*{\hsize}{!}{\includegraphics{multiopt4psr.ps}}
\resizebox*{\hsize}{!}{\includegraphics{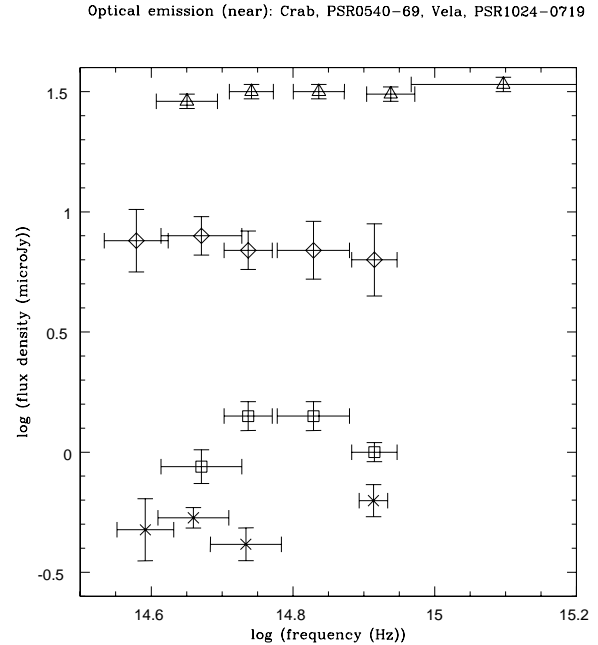}}
\end{figure}

The extinction-corrected magnitudes for 1024-Br, assuming it to be
at the same distance as the pulsar, are $U = 21.94$, $V = 19.71$,
$R = 18.81$, and $I = 18.12$. Similarly, the extinction-corrected
magnitudes for 1024-Fnt are $U = 23.65$, $V = 24.83$, $R = 24.33$
and $I = 24.19$. For the latter, the corresponding flux densities
(in $\mu\rm Jy$) at the frequencies corresponding to the centroids
of the FORS1 broadband Bessel filters (\cite{FORS1}) are plotted
in Fig. \ref{fig: multibandpsr}.

%          The extinction-corrected magnitudes for 1024-Fnt,
%assuming it to be at the same distance as the pulsar,
% (see Tab.
%\ref{tab: mag} for magnitudes without the reddening corrections)
%are $U = 23.65$, $V = 24.83$, $R = 24.33$ and $I = 24.19$.
%Similarly, the extinction-corrected magnitudes for 1024-Br are $U
%= 21.94$, $V = 19.71$, $R = 18.81$, and $I = 18.12$. We convert
%the magnitudes and their photometric errors for the faint object
%1024-Fnt to flux densities (in $\mu\rm Jy$) and errors and using
%the centroids and FWHM of the FORS1 broadband Bessel filters
%(\cite{FORS1}) plot these in Fig. \ref{fig: multibandpsr}.

          The corresponding extinction-corrected magnitude limits
for PSR J1744-1134 are $B \geq 26.8$ (corresponding to $0.079 \;
\mu\rm Jy$ at B-band centroid $7.0\times 10^{14}\; \rm Hz$), $V
\geq 26.2$, and $R \geq 26.0$.
%The absorption corrected flux
%limits in the B-band is then $0.079 \; \mu\rm Jy$ (at B-band
%centroid $7.0\times 10^{14}\; \rm Hz$).
%$0.163\; \mu\rm Jy$ (at V-band centroid
%$5.4\times 10^{14}\; \rm Hz$ and $0.130\; \mu\rm Jy$ (at R-band centroid
%$4.57\times 10^{14}\; \rm Hz$).

\subsection{Nature of the bright star close to PSR J1024-0719}
\label{sec: fieldstar}

Since there are two point-like optical sources close to the
nominal radio timing position of the pulsar, we examine their
photometric colours and the spectrum of the brighter one to
constrain their probable nature. The brighter star is in fact
closer to the radio position of the pulsar, yet it is improbable
that this is the optical counterpart.

The colours calculated for 1024-Br
from  the VLT observations, $V - R = 0.93$ and $V - I = 1.65$, do
not match the optical and IR colors of white dwarfs very well. On
the other hand, the visual colors V-R and V-I of 1024-Br can be
approximately accounted for by a K5-type dwarf (with $M \sim 0.7
M_{\odot}$, \cite{All00}).

\subsubsection{Spectroscopy of the brighter star 1024-Br}
\label{sec: spectrum}

\begin{figure}
\caption{ Spectrum of the brighter star 1024-Br near PSR
J1024-0719. The spectrum corresponds to a typical early K-type
dwarf star. \label{fig: dante_1024Br_spec}} \vskip 0.8cm
%\resizebox*{\hsize}{!}{\includegraphics{dante_1024Br_spec.ps}}
\resizebox*{\hsize}{!}{\includegraphics{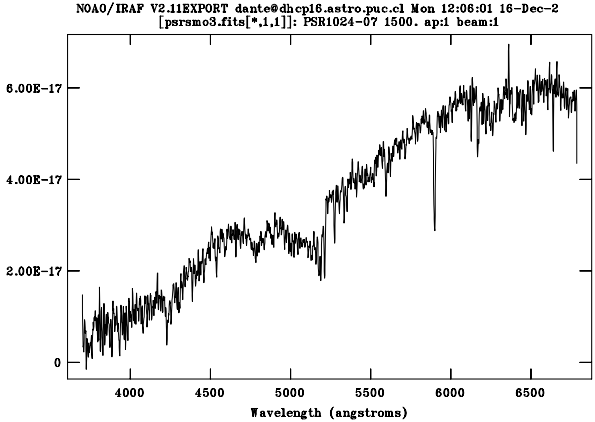}}
\end{figure}

Because of the proximity of 1024-Br to the radio position of the
pulsar (corrected for proper motion), spectroscopic observations
of this object were carried out. The spectra were acquired with
the Boller \& Chivens spectrograph on the Magellan I Baade 6.5 m
telescope at Las Campanas Observatory on the night of May 8, 2002.
These observations were taken during a spectroscopic survey of NGC
5128 globular clusters (see \cite{Minniti02} for more details on
the reduction procedures). We obtained two exposures of the source
1024-Br with 900 and 1500 seconds, respectively, at an airmass of
1.1. The measured seeing at the time of the observations was
$0\farcs8$, and a slit width of $1\farcs0$ was used. The
spectrophotometric standard LTT4816 observed at a similar airmass
was used for the flux calibration, although the latter is
uncertain due to the presence of thin cirrus during the night. The
spectral reductions and measurements were carried out in IRAF,
using the set of packages in CCDRED and TWODSPEC. The extracted
spectra cover from 3700 to $6780 \rm \AA$ , with $1.5 \rm \AA \;
pix^{-1}$. The final average spectrum corrected for cosmic rays
has a mean $S/N=70 \; \rm pix^{-1}$, and is shown in Fig.
\ref{fig: dante_1024Br_spec}. A few narrow absorption lines are
present, and the spectrum corresponds to a typical early K-type
dwarf star. Fig. \ref{fig: dante_1024Br_spec} shows absorption
features due to the Mg H band, the Mg5170 triplet, a weak Ca H and
K doublet, the Na D doublet, as well as a number of iron lines
such as Fe5270 and Fe5335. The lines of the hydrogen Balmer series
are weak, and no prominent emission lines are present that would
render this object peculiar in any way.

\subsubsection{Binary companion or background star?}
\label{sec: binarycomp}

One can nevertheless ask if the star 1024-Br is physically associated with
the pulsar in a possible binary system.
This is ruled out, considering that a K5 dwarf star is likely
to be as massive as $\approx 0.7 M_{\odot}$ and the constraint
from radio timing measurements of PSR J1024-0719.

A planet of mass $m$ orbiting a pulsar of mass $M_p$ in an orbit
of semi-major axis $a_p$ and period $P_y$  years produces a
periodic timing residual of semi-amplitude (see e.g.
\cite{Phinney92})
$${a_p \sin i\over c}= 1700 \; \mu{\rm s} {m\over M_\oplus}\left[P_y {1.4M_{\odot}\over M_p}\right]^{2/3}
\sin i.$$
%Thus, for example in 1 year orbits around a $1.4 M_{\odot}$ neutron star at
%an orbital inclination angle $i = 60$ deg, the Earth (mass $M_\oplus$)
%would produce a timing residual of 1.5 msec, the moon
%of $18\; \mu\rm sec$.
%and the asteroid Ceres $0.3\; \mu\rm sec$ (\cite{Phinney92}).
%Since \cite{Tos99} have already timed PSR1024-07 for
%2-3 years, the $P_y$ must be larger than this
%while their claimed residual in timing is better than 1 microsec.
Monitoring of PSR J1024-0719 for over 3 years yields a timing
residual smaller than $1\; \mu\rm s$ (\cite{Tos99}) which, for an
orbital period smaller than this time span, constrains the mass of
any companion to $m\sin i$ less than a thousandth of an Earth
mass. A small but somewhat less constraining limit is obtained by
allowing a longer orbital period ($P_{orb} = 2 \pi /\Omega_{orb}$)
and using the acceleration term
%
%and orbital period for any companion star would be
%larger than $P_y \geq 2-3$  yr.
%Hence companion masses (if any)
%of the order of m sin i of less than a thousandth
%of a earth mass are implied.
%
%A small but somewhat less constraining limit on the companion
%mass is obtained by assuming a companion
%in a wide binary around the pulsar and using the acceleration term
in a binary orbit, $(P_0 \Omega_{orb}^2 a_p \sin i /c) \cos
(\Omega_{\rm orb}t)$, being bounded by the measured (apparent)
period derivative of the pulsar spin $\dot P$. Here, $P_0$ is the
intrinsic spin period. With the amplitude of the sinusoidally
varying term and an orbital radius of 200 AU (corresponding to
$1\arcs$ in projection at $d=200$ pc), we get a companion mass,
$m\sin i<4 M_{\rm Jupiter}$. Thus, 1024-Br, which is much brighter
than even a Jupiter-like companion at this distance, is unlikely
to be the binary companion of PSR J1024-0719. It is most likely a
near-solar-mass background star located far beyond the 200 pc
upper limit for the pulsar distance obtained from the Shklovskii
effect.
% and the pulsar's
% spin down rate $\dot P$.

\subsection{The faint object 1024-Fnt}
\label{sec: candidate}

In the $M_V$ vs. $V-I$ plane, we compare 1024-Fnt with the cooling
track for a $0.6 M_{\odot}$ white dwarf (DA or non-DA) given in
Fig 6 of \cite{Bergeron95}. The $V-I$ colour is 0.73.
% as in Tab. \ref{tab: colours} above.
The $M_V$ is estimated by using $A_V = 0.11$ as above, and the
distance modulus $\mu = 5 \log (d/{\rm pc}) - 5 + A_V = 6.61$ for
a distance of 200 pc, so that 1024-Fnt at that distance would have
an absolute magnitude $M_V = 24.2 - 6.6 = 17.6$. This
colour-magnitude position is well below the cooling curve of a
$0.6 M_{\odot}$ white dwarf (both a DA or a non-DA, i.e. hydrogen
rich or non-hydrogen atmosphere WDs above a carbon core) as in
Wood's (1995) evolutionary model. The multi-band spectrum of this
object and its possible nature are discussed in the next section.

\subsection{Implications for thermal and non-thermal models of the neutron star}
\label{sec: modelimplications}

If we take a single blackbody spectral distribution over a whole
neutron star surface, the deepest (B-band) limit for PSR
J1744-1134 (with $B \geq 26.9$ and $A_B =0.07$) corresponds to a
temperature
$$T_{\infty} \leq  2.1\times 10^6 {\rm K}
\left(d/{0.36 \;{\rm kpc}}\over R_{\infty}/10\; {\rm km}\right)^{2},$$
where $T_{\infty} = T (1 - 2GM/Rc^2)^{1/2}$ and $R_{\infty} =
R/(1-2GM/Rc^2)^{1/2}$ are the redshifted temperature and radius.
We note that the X-ray detection (BT99) already constrains this
temperature to much lower values.

We plot the multi-band flux densities of the faint source near the
radio pulsar PSR J1024-0719, i.e. of 1024-Fnt as a function of the
photon frequency in Fig. \ref{fig: multibandpsr}. For comparison,
we also plot in this figure the corresponding multi-band flux
densities from the Crab, LMC pulsar PSR B0540-69, and Vela
(obtained from \cite{Per93}; \cite{Mid87} as restated by
\cite{Nas97}; and \cite{Nas97} respectively). The remarkable
similarity of the source 1024-Fnt with the multi-band flux
densities of these well-known pulsars is noteworthy. For
comparison in other wave-bands, the spectral (energy) index of
Crab, $\gamma$ ($I_{\nu} \propto \nu^{-\gamma}$) is consistent
with $\gamma = 0$ at optical frequencies and varies from $\gamma =
1.1$ in the $\gamma$-ray region via 0.7 in the hard X-rays to 0.5
in the soft X-rays. The Crab spectral index changes sign and goes
to $\gamma = -2$ in the far infrared and its radio frequency
component is a separate component from the higher energy part
(flux density decreasing with increasing frequency again) and has
$\gamma = 2.7$ (\cite{Lyne90}).

\begin{figure}
\caption{Multiwaveband emission from near PSR J1024-0719. Included
are the radio flux densities at 400, 600 and 1400 MHz; the optical
data (this paper) of 1024-Fnt; and the ROSAT HRI X-ray data. Two
illustrative power laws and a blackbody curve are shown. The
blackbody emission is from a heated region of the neutron star
polar cap of radius 60 m and temperature $2 \times 10^6$ K, at a
distance of 50 pc. \label{fig: xor1024}} \vskip 0.7cm
%\resizebox*{\hsize}{!}{\includegraphics{xor_revdataprog_psr1024_BB2e6Rcore60m50pc.ps}}
\resizebox*{\hsize}{!}{\includegraphics{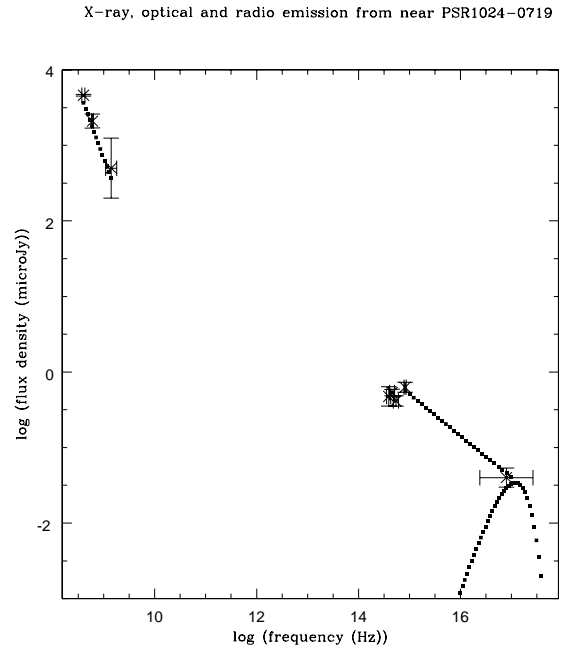}}
\end{figure}

       Similarly, the X-ray flux density of  PSR J1024-0719
in the ROSAT band and its standard error were computed from the
reported {\it un-absorbed} flux above (with the HRI count-rate)
and an assumed photon index $\alpha = 2.3$ ($\gamma = 1.3$), close
to that of the X-ray bright millisecond pulsar PSR J0437-4715 as
seen by ROSAT. This gives a flux density of $4.1 \times 10^{-2}
\mu\rm Jy$ with the centroid at 0.33 keV ($7.97\times 10^{16} \rm
Hz$) and an effective bandwidth of 0.23 keV. The radio flux
densities at three frequencies (400 MHz, 600 MHz and 1400 MHz) for
both pulsars were obtained from Bailes et al. (1997).
%The multi-band flux density spectrum for PSR J1024-0719 from
%radio to X-ray is plotted in Fig. \ref{} along with the optical
%magnitudes of the source 1024-Fnt from this (VLT) observation.
To guide the eye in Fig. \ref{fig: xor1024} for the multi-band
data, we also draw a few representative spectra: a) a power law
with photon index $\alpha_r = 2.85$  ($\gamma = 1.85$), which
matches the radio data well; b) a power-law spectrum of index
$\alpha_{ox} = 1.55$ connecting the highest frequency optical
datum to the X-ray, and c) a blackbody spectrum of temperature
$T_{\infty} = 2.1 \times 10^6$ K from a heated cap
($R_{\infty}^{core} = 0.06$ km) of the neutron star surface at a
distance of $d=0.2$ kpc. It is clear
%from Fig. \ref{fig: xor1024}
that a single power-law connecting the optical emission from
1024-Fnt and the X-ray emission from PSR J1024-0719 shall involve
a minimum photon index of 1.55 and the maximum limit on the hot
polar cap temperature $T \leq 2.1 \times 10^6 \rm K$ for a
hot-spot of radius 60 m. On the other hand, the optical magnitude
in the U-band implies for a single blackbody function for the
entire neutron star surface a temperature
\begin{eqnarray}
T_{\infty} & \leq & 4 \times 10^4 {\rm K} / \ln\{1 + 4.35
\times 10^{-4} [(R_{\infty}/10 {\rm km})/d_{\rm kpc}]^2\} \nonumber \\
   & \approx & 3.7 \times 10^6 {\rm K} \left(d/{0.2\;\rm kpc}\over
R_{\infty}/{10\;\rm km}\right)^2, \nonumber
\end{eqnarray}
where the latter expression corresponds to the Rayleigh-Jeans
limit. This limit is again much less constraining than that
obtained from the X-ray data of BT99.

\section{Discussion}
\label{sec: disc}

Non-thermal models for optical radiation for slower pulsars like
the Crab ascribe it to synchrotron emission (see \cite{Pac71} and
\cite{Pac87}) by relativistic particles near the light cylinder
radius at very small pitch angles ($\Psi \ll 1/\gamma$, where the
Lorentz factor may be $\gamma \sim 10^2 - 10^3$; \cite{Wat01}).
\cite{Mal01} show that finite pitch angles of the plasma particles
appear in the outer magnetosphere near the light cylinder due to
an instability associated with the cyclotron resonance condition
being fulfilled there. They further show that pulsars with shorter
periods such as the millisecond pulsars attain harder spectra and
higher peak frequencies. Although millisecond pulsars have much
smaller surface magnetic fields than slower pulsars, they also have
smaller corotating magnetosphere radii, making the dipolar fields
at light cylinder distances (a key parameter for particle
acceleration models) similar to those of slower pulsars. Thus, in
many of the millisecond pulsars, acceleration of particles in
outer gaps sustained by pair creation by GeV gamma-rays may still
be operative as in high spin-down luminosity pulsars
(\cite{Cheng86}). The non-thermal power-law component due to
pair-gap discharge is expected to scale with the Goldreich-Julian
particle flux $\dot N_{GJ}=({\bf\Omega\cdot B})/2 \pi e$ from the
magnetosphere (see \cite{Harding81} or \cite{Thompson98}). For
some millisecond pulsars $\dot N_{GJ}$ is in excess of that of PSR
B0656+14, which is found to be a gamma-ray, X-ray and optical
pulsar. We note that for our target pulsars, the Goldreich-Julian
particle fluxes are close to that of Geminga (within a factor of
two) from which optical pulsations have possibly been seen
(\cite{Shearer98}).

Scaled to the period ($P = 5.16$ ms) and the surface magnetic
field ($B_0 \sim 10^{8.1}$) in PSR J1024-0719, expressions (33)
and (35) of Crusius-W\"atzel et al. (2001) imply that

$$ {I_{\nu}^{opt} \over I_{\nu}^X } = 6 \times 10^{-2}
{(\Psi_{opt}/10^{-3})^2 (\nu_{opt}/10^{15}{\rm Hz}) (P / 5.16{\rm
\;ms})^{3/2} \over (\Psi_X/0.1)^{3/2} (\nu_X/10^{17}{\rm
Hz})^{1/2} (B_0 / 10^{8.1}{\rm G})^{1/2}}. $$ This ratio is
smaller than what is found for the optical to X-ray flux density
ratio of PSR J1024-0719 (see Fig. \ref{fig: xor1024}), i.e.,
$I^{opt} / I^X \sim 15$ instead of $6\times 10^{-2}$, if the
object 1024-Fnt is indeed the pulsar optical counterpart,
and the observed emission is primarily pulsed. This
indicates that in millisecond pulsars the average pitch angles
responsible for optical radiation are larger, $\Psi_{opt} \sim
10^{-2}$.

In middle-aged ($\tau \approx 10^5 - 10^6$ yr) and even older
neutron stars, at least a part of the high-energy radiation is due
to thermal emission from the neutron star's surface. The faint
measured UV flux from the slow pulsar PSR B0950+08 indicates that
a surface temperature $T_{\rm s} \sim 3 - 5\times 10^4 $ K could
be appropriate for an old pulsar, assuming a blackbody fit and
thermal radiation arising from the entire NS surface (Pavlov,
Stringfellow, \& Cordova 1996). Even more than for this relatively
old pulsar, the much older age of the millisecond pulsars implies
that these stars have almost certainly radiated away any
``fossil'' heat due to the original collapse or to subsequent
accretion from a binary companion. However, their high spin-down
power implies that internal dissipation is probably active,
keeping their surface at a higher temperature (e.g., \cite{Che92},
\cite{Rei95}, \cite{Rei97}, \cite{Lar99}, \cite{Sch99}).
Unfortunately, the optical band limits obtained here are not
strongly constraining for a blackbody of size similar to or
smaller than a NS.

The object 1024-Fnt has a multi-band spectrum which is unusual and
characteristic of pulsars, such as the Crab, Vela and PSR
B0540-69. Given its proximity to the radio position of PSR
J1024-0719 and the considerable astrometric uncertainty, only
timing analysis with high time resolution can unambiguously define
its nature.

\section{Conclusions}
\label{sec: conclusn} Among the isolated millisecond pulsars, none
so far have been shown to pulse in the optical. Deep exposures
carried out with the VLT of the field of the nearby, isolated
pulsar PSR J1744-1134 in multiple bands do not show a plausible
optical counterpart. We derive upper limits to the surface
temperature of the underlying neutron star at the likely distance
of the pulsar. In the field of view of PSR J1024-0719, we detect
two objects close to the radio position of the pulsar. We argue,
based on photometry and spectroscopy, that the brighter object is
a background early K-type dwarf star. The fainter object,
1024-Fnt, has a multi-band spectrum which is unusual and
characteristic of pulsars. Future high time resolution
observations may unambiguously define its nature by the detection
of optical pulses.

\begin{acknowledgements}
We thank M. Cristina Depassier for her role in the formation of
this collaboration and Marina Rejkuba for sharing with us the
Magellan I spectroscopic data obtained in collaboration with D.
Minniti. A. Ray thanks Malvin Ruderman for discussions on optical
radiation from pulsars at the Aspen Center for Physics and Poonam
Chandra for her comments on the manuscript. A. Reisenegger thanks
Marten van Kerkwijk for advice about astrometry with the VLT,
Fr\'ed\'eric Courbin for help with the ``p2pp'' and Ren\'e
M\'endez for useful discussions. A. Reisenegger's and G.
Hertling's work was supported by FONDECYT grant No. 1020840, and
H. Quintana's and D. Minniti's by the FONDAP Center for
Astrophysics. At Tata Institute, this work was a part of the Five
Year Plan Projects 10P-201 and 9P-208[a].
\end{acknowledgements}


\begin{thebibliography}{99}
\bibitem[Allen 2000]{All00} Allen C. W. 2000, ``Allen's Astrophysical Quantities'',
(4th. edition), Ed. A. N. Cox, Springer.

\bibitem[Alpar et al., 1982]{Alp82} Alpar M. A., Cheng A., Ruderman M.A., Shaham J.
1982, Nature, 300, 728.

\bibitem[Bailes et al. 1997]{Bailes97} Bailes M., Johnston S., et al.,
%Bell, J. F.,
%Lorimer, D. R., Stappers, B. W., Manchester, R. N., Lyne, A. G.,
%Nicastro, L., D'Amico, N., Gaensler, B. M. 1997,
ApJ 481, 386.

\bibitem[Becker \& Aschenbach 2002]{Beck02} Becker W., Aschenbach B. 2002,
in ``Neutron stars, pulsars and supernova remnants'', WE-Heraeus
Seminar, p.64.

%\bibitem[Becker \& Pavlov 2001]{Beck01} Becker W., Pavlov G. 2001,
%in ``Century of Space Science'', Eds. J. Bleeker et al., Kluwer
%Academic Publishers.

\bibitem[Becker \& Tr\"umper 1997]{Beck97} Becker W., Tr\"umper J. 1997 A\&A 326, 682.

\bibitem[Becker \& Tr\"umper 1999]{Beck99} Becker W., Tr\"umper J. 1999, A\&A 341,
803-817: BT99.

\bibitem[Bergeron et al. 1995]{Bergeron95} Bergeron P., Wesemael F., Beauchamp A. 1995, PASP 107, 1047.

%\bibitem[Bergeron et al. 2001]{Berg01} Bergeron P., Leggett S. K.,
%Ruiz M. T. 2001, ApJS, 133, 413.

%\bibitem[Bessel \& Brett 1988]{Bes88} Bessel M. S., Brett J. M. 1988, PASP 100, 1134.

\bibitem[de Boer et al. (1987)]{dB87}  de Boer K. S., et al. 1987, in ``Exploring
the Universe with IUE'', ed. Y. Kondo, D. Reidel and Company.

%\bibitem[Chen \& Ruderman 1993]{Chen93} Chen K., Ruderman
%M. 1993 ApJ 408, 179.

%\bibitem[Cheng et al. 2000]{Cheng00} Cheng K. S., Ruderman M.,
%Zhang L. 2000 ApJ 537, 964.

\bibitem[Cheng et al. 1992]{Che92} Cheng K. S., Chau W. Y., Zhang J. L.,
Chau H. F. 1992, ApJ 396, 135.

\bibitem[Cheng, Ho, \& Ruderman 1986]{Cheng86} Cheng K. S., Ho C.,
Ruderman M. A. 1986, ApJ 300, 500.

%\bibitem[Cocke, Disney, \& Taylor 1969]{Cocke69} Cocke W.J., Disney M.J.,
%Taylor D.J. 1969, Nature, 221, 525.

\bibitem[Crusius-W\"atzel et al. 2001]{Wat01} Crusius-W\"atzel A. R.,
Kunzl T., Lesch H. 2001, ApJ 546, 401.

\bibitem[Fomalont et al. (1992)]{Fom92} Fomalont E. B., Goss W. M.,
Manchester R. N., Lyne A. G., Justtanont K. 1992, MNRAS, 258, 497.

\bibitem[Fomalont et al. 1997]{Fom97} Fomalont E. B., Goss W. M.,
Manchester R. N., Lyne A. G. 1997, MNRAS, 286, 81.

\bibitem[FORS1 2000]{FORS1} FORS1+2 User Manual, Paranal Observatory
Very Large Telescope, European Southern Observatory Manual,
p. 87.

%\bibitem[Greenstein 1975]{Gre75} Greenstein G. 1975, ApJ 200, 281.

\bibitem[Harding 1981]{Harding81} Harding A. 1981, ApJ 245, 267.

\bibitem[Hewish et al. 1968]{Hew68} Hewish A., Bell S. J., Pilkington J.D.H.,
Scott P.F., Collins R.A. 1968, Nature, 217, 709..

%\bibitem[Iida \& Sato 1997]{Iid97} Iida K.,  Sato K. 1997, ApJ 477, 294.

\bibitem[Landolt (1992)]{Land92} Landolt A.U. 1992, AJ, 104, 340.

\bibitem[Larson \& Link 1999]{Lar99} Larson M. B., Link B. 1999,
ApJ 521, 271.

\bibitem[Lyne \& Smith 1990]{Lyne90} Lyne A. G., Graham-Smith F. 1990,
``Pulsar astronomy'', Cambridge Univ. Press.

\bibitem[Ma et al. 1998]{Ma98} Ma C. et al. 1998, AJ 116, 516.

\bibitem[Malov \& Machabeli (2001)]{Mal01} Malov I., Machabeli G.Z. 2001,
ApJ 554, 587.

\bibitem[Massey \& Davis 1992]{Massey} Massey P., Davis L. E. 1992,
``A Users guide to stellar photometry IRAF''.

\bibitem[Mathis 1999]{Allen99} Mathis J. 1999, ``Allen's Astrophysical
Quantities'', A. N. Cox (Editor), Springer-Verlag.

\bibitem[Minniti \& Rejkuba 2002]{Minniti02} Minniti D., Rejkuba M. 2002, ApJ 575, L59.

\bibitem[Middleditch et al. 1987]{Mid87} Middleditch J., Pennypacker C.R.
Burns M.S. 1987, ApJ 315, 142.

%\bibitem[Monet et al. 1998]{Mon98} Monet D.
%     Bird A., Canzian B., Dahn C., Guetter H., Harris H., Henden A.,
%     Levine S., Luginbuhl C., Monet A.K.B., Rhodes A., Riepe B.,
%     Sell S., Stone R., Vrba F.,  Walker R. 1998, Vizier On-line
%Data Catalog: I/252. U.S. Naval Observatory Flagstaff Station (USNOFS)
%and, Universities Space Research Association (USRA) stationed at USNOFS.

\bibitem[Monet et al. 1998]{Mon98} Monet D., Bird A., %Canzian B. et al.,
et al.,
 Vizier On-line Data Catalog: I/252. U.S. Naval Observatory Flagstaff
 Station (USNOFS).
%and Universities Space Research Association (USRA)
%stationed at USNOFS.

\bibitem[Nasuti et al. 1997]{Nas97} Nasuti F.P., Mignani R.,
Caraveo P.A., Bignami G.F. 1997, A\&A 323, 839.

\bibitem[Pacini 1971]{Pac71} Pacini F. 1971, ApJ 163, L17.

\bibitem[Pacini \& Salvati 1987]{Pac87} Pacini F., Salvati M. 1987,
ApJ 321, 447.

\bibitem[Pavlov et al. 1996]{Pav96} Pavlov G. G., Stringfellow G. S.,
 Cordova F. A. 1996, ApJ 467, 370.

\bibitem[Percival et al. 1993]{Per93} Percival J.W., et al. 1993, ApJ 407, 276.

\bibitem[Phinney 1992]{Phinney92} Phinney E. S. 1992, Phil. Trans. Roy. Soc.
Lond. A 341, 39.

\bibitem[Reisenegger 1995]{Rei95} Reisenegger A. 1995, ApJ 442, 749.

\bibitem[Reisenegger 1997]{Rei97} Reisenegger A. 1997, ApJ 485, 313.

\bibitem[Rots et al. 1997]{Rots97} Rots A. H., Jahoda K., Macomb D., et al.
%,  Kawai N., Saito Y., Kaspi V. M., Lyne A. G., Manchester R. N.,
%Backer D. C., Somer A. L., Marsden D.,  Rothschild R. E.
1998 ApJ 501, 749.

\bibitem[Ruderman \& Sutherland 1975]{Ruderman75} Ruderman M. A., Sutherland P. G.
1975, ApJ 196, 51.

\bibitem[Ruiz \& Bergeron 2001]{Ruiz01} Ruiz M. T., Bergeron P. 2001,
ApJ 558, 761.

\bibitem[Schaab et al. 1999]{Sch99} Schaab Ch., Sedrakian A., Weber F.,
Weigel M. K. 1999, A\&A, 346, 465.

\bibitem[Shklovskii 1970]{Shk70} Shklovskii I. S. 1970, Soviet
Astron., 13, 562.

\bibitem[Shearer \& Golden 2002]{Shearer02} Shearer A., Golden A. 2002,
in ``Neutron stars, Pulsars and Supernova Remnants'', WE-Heraeus
Seminar, Eds. W. Becker et al., MPE Report 278, p. 44.

\bibitem[Shearer \& Golden 2001]{Shearer01} Shearer A., Golden A. 2001,
ApJ 547, 967.

\bibitem[Shearer et al. 1998]{Shearer98} Shearer A., Golden A., et al. 1998,
A\&A 335, L21.

\bibitem[Staelin \& Reifenstein, 1968]{Staelin68} Staelin D. H.,
Reifenstein E. C. 1968, Science, 162, 1481.

\bibitem[Taylor \& Cordes (1993)]{Tay93} Taylor J. H.,  Cordes J. M. 1993,
ApJ 411, 674.

\bibitem[Thompson 1998]{Thompson98} Thompson D. J. 1998, in ``Neutron
stars and pulsars'', ed. N. Shibazaki et al., Universal Academic
Press (Tokyo).

\bibitem[Toscano et al. 1999a]{Tos99} Toscano M., Britton M. C.,
Manchester R. N., et al. 1999a,
%Bailes M., Sandhu J. S., Kulkarni S. R.,
%Anderson S. B., Stappers, B. W. 1999,
MNRAS, 307, 925.

\bibitem[Toscano et al. 1999b]{Tos99_2} Toscano M., Britton M. C.,
Manchester R. N., et al.,
%Bailes, M., Sandhu, J. S., Kulkarni, S. R.,  Anderson, S. B.
1999b, ApJ 523, L171.

\end{thebibliography}
\end{document}